\documentclass[epj]{svjour}
\usepackage{amsmath,amssymb}
\usepackage{graphics}
\usepackage{epsfig}

\begin{document}
\title{Andreev Bound States in High Temperature Superconductors}
\author{L.~Alff \and S.~Kleefisch \and U.~Schoop \and M.~Zittartz
\and T.~Kemen \and T.~Bauch \and A.~Marx \and R.~Gross}
\institute{II.~Physikalisches Institut, Universit\"{a}t zu K\"{o}ln,
Z\"{u}lpicherstr.~77, D - 50937 K\"{o}ln, Germany}

\date{Received: March 17, 1998}

\abstract{Andreev bound states at the surface of superconductors are
expected for any pair potential showing a sign change in different
$k$-directions with their spectral weight depending on the relative
orientation of the surface and the pair potential. We report on the
observation of Andreev bound states in high temperature superconductors (HTS)
employing tunneling spectroscopy on bicrystal grain boundary Josephson
junctions (GBJs). The tunneling spectra were studied as a function of
temperature and applied magnetic field. The tunneling spectra of GBJ formed
by YBa$_2$Cu$_3$O$_{7-\delta}$ (YBCO), Bi$_2$Sr$_2$CaCu$_2$O$_{8+\delta}$
(BSCCO), and La$_{1.85}$Sr$_{0.15}$CuO$_4$ (LSCO) show a pronounced zero bias
conductance peak that can be interpreted in terms of Andreev bound states at
zero energy that are expected at the surface of HTS having a $d$-wave
symmetry of the order parameter. In contrast, for the most likely $s$-wave
HTS Nd$_{1.85}$Ce$_{0.15}$CuO$_{4-y}$ (NCCO) no zero bias conductance peak
was observed. Applying a magnetic field results in a shift of spectral weight
from zero to finite energy. This shift is found to depend nonlinearly on the
applied magnetic field. Further consequences of the Andreev bound states are
discussed and experimental evidence for anomalous Meissner currents is
presented.
\\ {\em Dedicated to J. Zittartz on the occasion of his 60th birthday}
}

\maketitle
\section{Introduction}
\label{intro}

Recent experimental and theoretical work on the symmetry of the order
parameter in the high temperature superconductors (HTS) has led to the
conclusion that for the majority of the cuprate superconductors the symmetry
of the order parameter is dominated by a $d_{x^2-y^2}$-wave component
\cite{vHarlingen:95,Scalapino:95}. This is in clear contrast to the metallic
low-temperature superconductors for which the order parameter is dominated by
an isotropic $s$-wave component. The $s$-wave order parameter usually has the
full symmetry of the point group of the underlying crystal structure and,
hence, at the transition to the superconducting state only the global gauge
symmetry is broken. Such pairing symmetry is denoted conventional. In
contrast, the $d$-wave order parameter of the cuprate superconductors may not
have the full symmetry of the point group of the crystal structure and,
hence, beyond the global gauge symmetry a further symmetry is broken at the
transition to the superconducting state. Such pairing symmetry then is
denoted unconventional. An important characteristic of the
$d_{x^2-y^2}$-symmetry of the order parameter prevailing in most HTS is a
sign change of the pair potential in orthogonal $k$-directions what is
equivalent to a $\pi$-phase shift of the wave function. Usually the positive
sign is taken along the $a$-axis of the unit cell and the negative sign along
the $b$-axis. Furthermore, there are nodes of the pair potential in the [110]
directions, where it changes sign. That is, the quasiparticles with
corresponding $k$-vector actually feel a vanishing pair potential what allows
for low energy quasiparticle excitations. This is in contrast to a $s$-wave
symmetry, where a finite gap on the whole Fermi surface does not allow for
quasiparticle excitations of arbitrarily small energy.  We note that the
actual order parameter present in hole doped HTS likely is formed by a
mixture of a dominating $d_{x^2-y^2}$-component and other components such as
an $s$- or $d_{xy}$-component. Furthermore, there is evidence that the order
parameter of the the electron doped HTS Nd$_{1.85}$Ce$_{0.15}$CuO$_{4-y}$
(NCCO) has a dominating $s$-wave component and no sign change in different
$k$-directions.

There have been numerous experiments devoted to the determination of the
symmetry of the order parameter in the oxide superconductors. These
experiments can be devided into those probing the amplitude by studying the
quasiparticle excitation spectrum and those probing the phase of the order
parameter in interferometer experiments employing multiply connected
superconductors. Certainly, the key experiments have been the phase sensitive
experiments designed by Tsuei and Kirtley which employ HTS thin film
tricrystals \cite{Kirtley:95,Tsuei:96}. In these experiments a scanning
SQUID-system has been used to show that there are half-integer flux quanta in
superconducting rings formed by three differently oriented HTS grains
connected by grain boundary Josephson junctions. In this article we present a
new class of phase sensitive experiments making use of the formation of
Andreev bound states at surfaces of HTS oriented parallel to the $c$-axis. An
important consequence of an order parameter showing a sign change (or $\pi$
phase shift) in different $k$-directions, is the formation of Andreev bound
states at zero energy confined to the surface. Andreev bound states have been
discussed first in the context of tunneling into unconventional
superconductors by Buchholtz and Zwicknagel \cite{Buchholtz:81}. Later on, the
formation of Andreev bound states or midgap surface states having zero energy
with respect to the Fermi energy and sizable areal density as a consequence of
a $d_{x^2-y^2}$-symmetry of the order parameter has been predicted by Hu
\cite{Hu:94}. These zero energy Andreev bound states can be probed by
$ab$-plane tunneling spectroscopy and manifest themselves as a zero bias
conductance peak. That is, the presence of Andreev bound states at zero energy
represents definite evidence that the order parameter of the investigated
superconductor changes sign along the Fermi surface. We emphasize that the
same $\pi $ phase shift in the Josephson interference experiments by Tsuei
{\it et al.} \cite{Tsuei:96} is the origin of the Andreev bound states and the
zero bias conductance peak  in the $ab$-plane tunneling conductance.

In this article we present comprehensive experimental data on $ab$-plane
tunneling spectroscopy for various HTS materials. The data have been obtained
using [001] tilt HTS grain boundary Josephson junctions (GBJs). In contrast
to low temperature scanning tunneling spectroscopy and experiments using
SIN-type planar junctions, GBJs represent
superconductor-insulator-super\-con\-duc\-tor (SIS) Josephson junctions
\cite{Gross:94a,Gross:twente,Gross:97a}. An important advantage of the use of
GBJs as compared to low temperature scanning tunneling spectroscopy is the
very good long term stability of GBJs which allows for the detailed study of
the temperature and magnetic field dependence of the tunneling spectra.
Furthermore, in GBJs one deals with internal surfaces or interfaces. This
implies that degradation or contamination effects due to {\em ex situ}
processing in ambient atmosphere are of minor importance. Only oxygen
depletion in the region of the grain boundary can occur, but to a less extent
as compared to bare surfaces used in low temperature scanning tunneling
spectroscopy experiments.  The tunneling data of the recent experiments using
SIN-type junctions could be consistently explained by the formation of
Andreev bound states as a consequence of a dominating $d$-wave symmetry of
the order parameter in the investigated HTS materials
\cite{Kashiwaya:95a,Alff:97b,Covington:96,Covington:97}. Here, we show that
the same is true for the SIS-type GBJs fabricated from different HTS
materials. Our experimental results can be compared to recent theoretical
predictions by Barash {\it et al.}~\cite{Barash:95a} and Tanaka {\it et
al.}~\cite{Tanaka:97n} on the Josephson behaviour of junctions with $d$-wave
electrodes. Our detailed experimental study clearly shows that the zero bias
conductance peak observed in HTS-GBJs is caused by Andreev bound states.
Competing explanations for the origin of the zero bias conductance peak, in
particular the magnetic scattering scenario, which is based on a model by
Anderson and Appelbaum \cite{Appelbaum:66,Anderson:66} developed for NIN-type
junctions containing magnetic impurity states, can be ruled out.

Our analysis includes tunneling data obtained for the three hole doped HTS
materials YBa$_2$Cu$_3$O$_{7-\delta}$ (YBCO),
Bi$_2$Sr$_2$CaCu$_2$O$_{8+\delta}$ (BSCCO), and La$_{2-x}$Sr$_x$CuO$_4$
(LSCO). Here, YBCO has been investigated in the optimum doped phase with a
critical temperature $T_c$ of 90\ K and in the underdoped phase with a $T_c$
of about 60\,K.  Furthermore, the electron doped material NCCO has been
studied.  The analysis of this material is of particular interest with
respect to the question whether or not there is a change in sign of the order
parameter for NCCO. Up to now, there is convincing experimental evidence that
NCCO has a dominating $s$-wave symmetry of the order parameter
\cite{Huang:90,Wu:93,Andreone:94,Alff:96b}, i.e., the order parameter of NCCO
is expected to show no change in sign for different $k$-directions. Hence,
Andreev bound states are expected only for hole doped HTS, which are believed
to have a dominating $d$-wave symmetry of the order parameter, but should be
absent for NCCO. As we shall see below, this is indeed the case
\cite{Alff:96b,Ekin:97}.

\section{Theoretical background}
\label{TheoBack}

There are several theoretical models that can account for a zero bias
conductance peak  in the tunneling spectra of SIN- or SIS-type junctions
\cite{Hu:98}. Beyond the ABS scenario, the Appelbaum-Anderson model and the
Blonder-Tinkham-Klapwijk model, phase diffusion in HTS Josephson junctions
has been discussed \cite{Wilkins:90,Walsh:91}. However, any conductance peak
related to a supercurrent should be restricted to a much smaller voltage
scale than the mV-scale usually measured for the zero bias conductance peak
in most experiments. Moreover, any model based on supercurrents fails for
SIN-type junctions.  In the following we therefore will restrict our
discussion to the Appelbaum-Anderson model, , the Andreev bound state model,
and the Blonder-Tinkham-Klapwijk model.

\subsection{Appelbaum-Anderson model}
\label{AA}

Zero-bias anomalies in the conductance of normal metal-\-insulator-\-normal
metal (NIN) tunnel junctions often have been observed and can have various
reasons \cite{Shen:68,Wolf:book}. In a model developed by Appelbaum and
Anderson  it is assumed, that the tunneling barrier is associated with
localized paramagnetic states \cite{Appelbaum:66,Anderson:66}. The tunneling
conduction electrons are exchange coupled to these states leading to exchange
scattering off the localized states. Similar to the $s-d$ interaction model
giving a resistance minimum in dilute magnetic alloys (Kondo-type scattering
\cite{Kondo:64}), the Appelbaum-Anderson model can account for a zero bias
conductance peak. The tunneling conductance $G$ in the Appelbaum-Anderson
model is given by three terms $G=G_1+G_2+G_3$, where $G_1$ is the
contribution from all tunneling processes without spin interaction. $G_2$ is
the spin exchange process contribution to the conductance that, of course,
depends on the applied magnetic field but not on the voltage at zero field.
$G_3$ is the Kondo-type contribution where an electron is scattered by the
exchange interaction leading to the interference of reflected and transmitted
waves. This additional tunneling channel contributes logarithmically to the
conductance as

\begin{equation}\label{aa}
G_3(V,T)\propto\ln\left(\displaystyle\frac{E_0}{|eV| + nk_BT}\right).
\end{equation}

\noindent
Here, $E_0$ is an energy cut-off and $n$ a factor close to unity. An applied
magnetic field $H$ results in a Zeeman splitting of the impurtiy states. Then,
$G_2$ is strongly suppressed for quasiparticle energies $eV$ smaller than the
distance between the Zeeman-levels and rises abruptly for $|eV|\geq g\mu_BH$,
where $g$ is the Land\'{e} $g$-factor, and $\mu _B$ the Bohr magneton.
Furthermore, $G_3$ is splitted by $2\delta=2g\mu_BH$. Strictly speaking, $G_3$
splits into three peaks with one peak centered at zero-bias. However, the
spectral weight of this peak goes to zero as $\delta$ is increased, probably
impeding the measurement of this zero-energy peak \cite{Appelbaum:66}.

The Appelbaum-Anderson model has been successfully applied to Josephson tunnel
junctions formed by low-tem\-perature superconductors employing hydrogenated
amorphous silicon barriers \cite{Kroger:85}. Initially, it also was used to
explain the experimentally observed zero bias conductance peak in tunnel
junctions employing HTS electrodes
\cite{Covington:96,Walsh:92,Lesueur:92,Cucolo:93,Kashiwaya:94,Froehlich:97}.
However, as will be shown below the Appel\-baum-Anderson model cannot account
for the temperature and magnetic field dependence of the zero bias conductance
peak observed for the HTS junctions including the GBJs. We emphasize that in
general localized magnetic states in superconductors can lead to bound states
within the gap as pointed out long ago by Zittartz {\it et al.}
\cite{Zittartz:69}.

\subsection{Andreev bound state model}
\label{ABS}

The formation of Andreev bound states or midgap surface states in high
temperature superconductors having zero energy with respect to the Fermi
energy and sizable areal density as a consequence of a $d_{x^2-y^2}$-symmetry
of the order parameter has been predicted by Hu \cite{Hu:94}. The spectral
weight of these bound states has a maximum for (110) oriented surfaces,
whereas no such states are expected for (100) or (010) surfaces. That is, the
mid gap states exist for any specular surface except for the lobe direction of
the $d_{x^2-y^2}$-gap perpendicular to the surface. The physical reason for
the midgap states is the fact that quasiparticles reflecting from the surface
experience a change in the sign of the order parameter along their classical
trajectory and subsequently undergo Andreev reflection. Constructive
interference of incident and Andreev reflected quasiparticles result in bound
states confined to the surface. That is, the midgap states can be understood
in terms of Andreev reflections, where the quasiparticle changes from
particle-like to hole-like and {\em vice versa}, and {\bf k} changes sign
\cite{Andreev:64,Rainer:97}. Therefore, the midgap states also are denoted as
Andreev bound states. Reversal of the velocity by Andreev reflection always is
accompanied by a change in sign of the charge. Consequently, Andreev bound
states can carry current, where current conservation is maintained by
conversion of the bound state current to supercurrent far away from the
surface. The Andreev bound states result in a zero bias conductance peak  in
an $ab$-plane tunneling spectrum. Thus, methods that probe the quasiparticle
current on a surface or through an interface represent phase sensitive methods
in the sense that the existence of Andreev bound states gives definite
information on a change in sign of the pair potential. Hence, tunneling
spectroscopy becomes a valuable phase sensitive technique for probing the
symmetry of the order parameter in HTS.

%%%%%%%%%%%%%%%%%%%%%%%%  FIGURE 1 %%%%%%%%%%%%%%%%%%%%%%%%%

\begin{figure}[t]
\noindent
\vspace*{0cm}\\
\epsfig{file=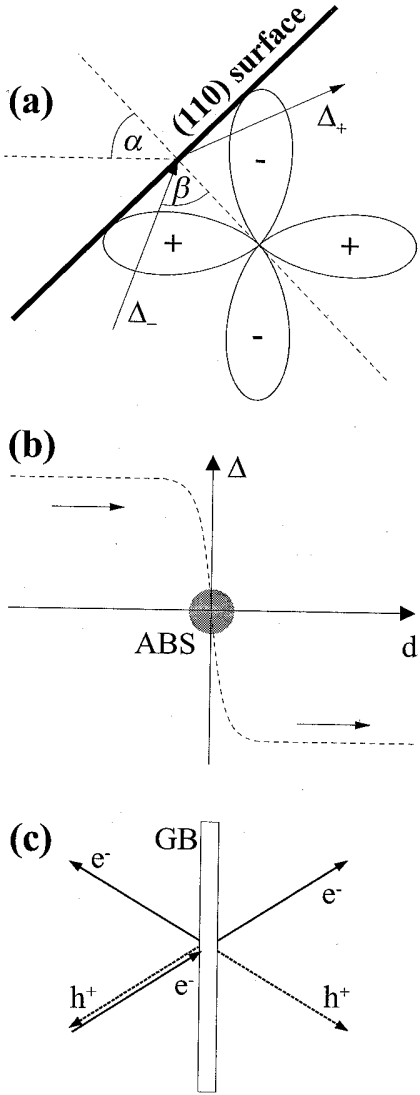,width=6.75cm,bbllx=0,bblly=550,bburx=195,bbury=1000,clip=}
\vspace*{-0.5cm}
\caption{Formation of Andreev bound states: In (a) a mixed real/momentum space representation
is shown for a (110) oriented surface of a HTS having a $d_{x^2-y^2}$
symmetry of the order parameter. The propagation direction of the
quasiparticle is indicated by the arrows and can be transferred to a
direction in $k$-space by plotting a $\vec{k}$-vector from the origin of the
lobe to its edge parallel to the propagation direction in real space. In (b)
the order parameter $\Delta$ is plotted versus the path $d$ along the
quasiparticle trajectory. For the case of a sign change on reflection the
Andreev bound states are always formed at the surface ($d=0$). In (c) a
schematic illustration of the different reflection and transmission processes
of quasiparticles at a grain boundary (GB) interface is given.}
\label{Fig:abs}
\end{figure}

%%%%%%%%%%%%%%%%%%%%%%%%%%%%%%%%%%%%%%%%%%%%%%%%%%%%%%%%%%%%

The basic mechanism required to understand the formation of surface or
interface bound states in $d_{x^2-y^2}$-HTS is the process of  Andreev
reflection \cite{Andreev:64}. In general, Andreev reflection always plays a
role if there is a variation of the order parameter along the classical
trajectory of a quasiparticle. That is, Andreev reflection can be viewed as
scattering induced by spatial variations of the order parameter. Such
variations of the order parameter are well known for interfaces in artificial
layer structures such as NS-interfaces. However, they also can occur along the
trajectory of quasiparticles specularly reflected at the surface of a
superconductor having a spatially anisotropic order parameter. In the process
of Andreev reflection a particle-like excitation undergoes branch conversion
into a hole-like excitation with reversed group velocity and {\em vice versa}.
In order to discuss the importance of Andreev reflection for the formation of
Andreev bound states at surfaces and interfaces of HTS in more detail let us
consider the situation shown by Fig.~\ref{Fig:abs}(a). An incident
quasiparticle with a $k$-vector corresponding to the negative lobe of the pair
potential reflects specularly off the surface. For simplicity, a (110) surface
($\alpha = \pi/4$) and a quasiparticle that is incident under an angle $\beta$
is considered. The order parameter $\Delta(\vec{k},\vec{r})$ experienced by
the quasiparticle is a function of momentum $\vec{k}$ and space $\vec{r}$. It
is evident that incident and reflected wave packets propagate through
different order parameter fields.  For the situation shown in
Fig.~\ref{Fig:abs}(a) the order parameter has different sign for the incident
and reflected wave packet. This is shown more clearly in
Fig.~\ref{Fig:abs}(b), where we have plotted the order parameter field along
the classical trajectory of the quasiparticle. The rounding of $\Delta$ close
to the surface is due to pair-breaking effects and plays no role for the
formation of Andreev bound states. However, it becomes important if pair
breaking frees significant spectral weight for subdominant pairing channels as
will be discussed in section~\ref{Observe}.

In general, Andreev bound states occur at energies for which the phases of
Andreev reflected particle-like and hole-like excitations interfere
constructively. According to the Atiyah-Singer index theorem zero energy
bound states are always formed, if the scattering induces a change in sign of
the order parameter along the classical trajectory \cite{Atiyah:75}. For the
situation shown in Fig.~\ref{Fig:abs} this is always directly at the
interface ($d=0$). Thus, changing the incident angle $\beta $  will not
change the location of the Andreev bound states at $d=0$. Reducing $\alpha $
from $\pi /4 $ towards $\alpha =0$ results in a increasingly smaller range of
incident angles $\beta $ for which a change in sign of the order parameter
field is obtained. Thus, the spectral weight of the zero energy bound states
decreases with decreasing $\alpha $ and vanishes for $\alpha =0$, i.e. for a
(100) or (010) oriented surface.  That is, the spectral weight of the Andreev
bound states is largest for (110) surfaces and decreases continously towards
zero for (100) or (010) surfaces. It is evident that for trajectories
orthogonal to the CuO$_2$-plane no Andreev bound states are formed.

Above we have considered the situation present at the surface of a HTS. We
now briefly discuss the situation for a SIS-structure present for the GBJs
studied in our experiments. In Fig.~\ref{Fig:abs}(c) the possible
transmission and reflection processes at a grain boundary interface are shown
schematically. An incident particle-like excitation is injected from the left
bulk $d$-wave HTS under a finite angle with respect to the grain boundary
plane. Discussing the reflection processes, the particle-like excitation can
be either specularly or Andreev reflected. In the latter case it is turned
into a hole-like excitation propagating in opposite direction. In the same
way, transmission through the insulating grain boundary barrier can yield
particle- and hole-like excitations. As described above for the case of
reflection at a surface, the quasiparticles involved in the different
reflection and transmission processes experience different pair potentials
depending on their $k$-vector and the relative orientation of the order
parameter in both electrodes with respect to the grain boundary barrier in
both electrodes. We also note that Cooper-pairs have to be involved to
establish charge conservation.

We point out that different order parameter symmetries will produce different
$k$-dependencies of the spectral weight of Andreev bound states. Of course,
Andreev bound states are present only if there is a change in sign of the pair
potential in different $k$-directions and are absent if there is none. For
example, for a $d_{xy}$- and $d_{x^2-y^2}$-symmetry the maximum spectral
weight of the Andreev bound states is obtained for a (100) or (110) oriented
surface or interface, respectively. In contrast, for a $s$-wave symmetry no
sign change of the order parameter and, hence, no Andreev bound states are
present. Thus, the experimental observation of Andreev bound states gives
definite evidence for an order parameter changing sign on the Fermi surface
and represents a phase sensitive probing technique. In cases where the order
parameter symmetry is not fully established, as for example for the electron
doped NCCO, Andreev bound states sensitive experiments can be used to clarify
whether or not the order parameter changes sign in different $k$-directions.

\subsection{Consequences of the Andreev bound state model}
\label{Observe}

There are several new and interesting phenomena that are directly related to
the existence of Andreev bound states at surfaces and interfaces of HTS that
will be addressed in the following. So far, there is significant experimental
evidence only for the zero bias conductance peak  in tunneling data due to
Andreev bound states. However, other effects like an anomalous temperature
dependence of the critical current of HTS Josephson junctions or an anomalous
temperature dependence of the London penetration depth due to
``anti-Meissner'' currents have not yet been experimentally confirmed.

\subsubsection{Zero bias conductance peak}

Since Andreev bound states can carry current the most evident experimental
consequence is that they should produce a pronounced zero bias conductance
peak  in the $ab$-plane tunneling conduction of experiments involving at least
one $d$-wave HTS electrode. For the interpretation of the conductance of
super\-con\-ductor-insulator-normal metal (SIN) junctions that can be formed
by using a normal metal tip in low temperature scanning tunneling
spectroscopy, several groups extended the calculations by Hu \cite{Hu:94}. In
particular, expressions for the density of states and the conductance of such
junctions employing $d$-wave HTS have been derived
\cite{Barash:95a,Tanaka:97n,Tanaka:95b,Buchholtz:95,Kashiwaya:96a,Barash:97}.
The most prominent feature in the calculated conductance versus voltage curves
indeed was a pronounced zero bias conductance peak  as a direct consequence of
the Andreev bound states. Recently, these states also have been observed
experimentally using low temperature scanning tunneling spectroscopy
\cite{Kashiwaya:95a,Koyanagi:92,Chen:92}, in particular on well defined (110)
oriented HTS surfaces \cite{Alff:97b}. We note that Andreev bound states are
absent at surfaces perpendicular to the $c$-axis direction. Therefore, no zero
bias conductance peak  has been observed by Renner {\it et al.} in low
temperature scanning tunneling spectroscopy experiments performed on (001)
oriented surfaces \cite{Renner:95,Renner:96,Renner:98}. A further possibility
for the realization of SIN-junctions is the formation of planar junctions
using a normal metal and a HTS electrode separated by a thin insulating layer.
Also for such junctions a zero bias conductance peak  has been observed in
most experiments
\cite{Covington:96,Covington:97,Walsh:92,Lesueur:92,Cucolo:93,Sinha:98}. Up to
now the theoretically predicted zero bias conductance peak  has been found for
both SIN- and SIS-type junctions
\cite{Walsh:91,Mandrus:91,Iguchi:92,Becherer:93}. Below we present
comprehensive experimental data on the tunneling conductance of various high
temperature superconducting GBJs giving a very complete view of the zero bias
conductance peak  as a consequence of Andreev bound states. We note that non
zero-energy states are also possible as a consequence of a suppression of the
order parameter at the surface \cite{Barash:97}.

\subsubsection{Surface roughness}

\begin{figure}[t]
\noindent
\vspace*{0.25cm}\\
\epsfig{file=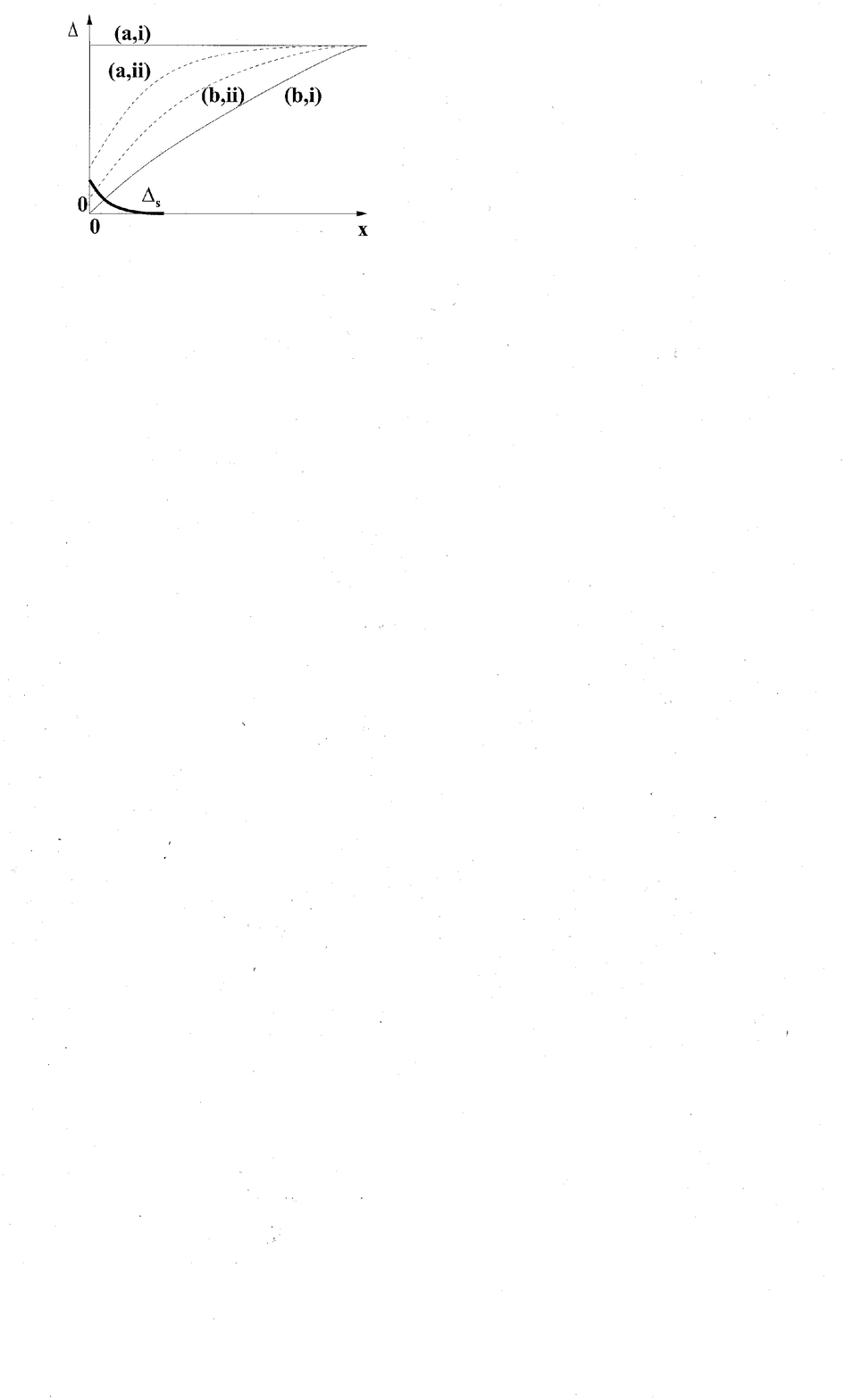,width=7.5cm,bbllx=40,bblly=820,bburx=285,bbury=1000,clip=}
\vspace*{-0.0cm}
\caption{Qualitative plot of the amplitude of the order parameter as a function of
the distance from the surface for a (100) (a) and  (110) (b) surface without
(i) and with (ii) surface roughness, respectively. Also shown is a surface
induced $s$-wave order parameter that can be obtained for a (110) surface by
surface pair breaking effects for the dominant $d$-wave and subsequent
pairing into the subdominant $s$-wave channel.}
\label{Fig:rough}
\end{figure}

In our discussion of the Andreev bound state model we always have assumed
perfectly flat surfaces and interfaces that usually are not present in any
experimental situation. The spectral weight of the Andreev bound states and,
hence, the measured zero bias conductance peak  is very sensitive to surface
roughness on a length scale larger than the coherence length. Since the
coherence length of HTS is of the order of 1\,nm, almost all surfaces can be
considered rough. The influence of surface roughness has been theoretically
calculated by several authors
\cite{Barash:97,Matsumoto:95a,Yamada:96,Tanuma:96}. The basic consequence of
surface roughness is that one has to average over different surface
orientations. As a consequence, for a $d_{x^2-y^2}$-symmetry of the order
parameter and a (110) oriented surface, for which the spectral weight of the
Andreev bound states is maximum, the zero bias conductance peak  is
suppressed by surface roughness. {\em Vice versa}, a finite zero bias
conductance peak  can appear for (100), (010), and (001) surfaces and
interfaces due to surface roughness. That is, surface roughness smears out
the pronounced dependence of the Andreev bound states spectral weight on the
relative orientation of the surface and the order parameter. As an example,
in Fig.~\ref{Fig:rough} the order parameter is plotted versus the distance
from the surface for smooth and rough surfaces with different orientation. It
is evident that due to surface roughness differently oriented surfaces behave
in a similar way. This makes it experimentally more difficult to distinguish
the different directions within the superconducting plane. However, since the
sign change of the pair potential is a necessary prerequisite for the
formation of Andreev bound states, any amount of surface roughness can not
produce a zero bias conductance peak  for an order parameter having no sign
change on the Fermi surface such as an $s$-wave order parameter.

%%%%%%%%%%%%%%%%%%%%%%%%  FIGURE 2 %%%%%%%%%%%%%%%%%%%%%%%%%

%%%%%%%%%%%%%%%%%%%%%%%%%%%%%%%%%%%%%%%%%%%%%%%%%%%%%%%%%%%%

\subsubsection{Anomalous temperature dependence of the Josephson current density}

A further important consequence of the Andreev bound states is an anomalous
behaviour of the temperature dependence of the maximum Josephson current $I_c$
at low temperatures \cite{Tanaka:96c,Barash:96b}. Recently, is was predicted
by Tanaka and Kashiwaya \cite{Tanaka:96c} that for a fixed phase difference
between the electrodes of a Josephson junction formed by $d$-wave
superconductors the supercurrent across the junction can be either positive or
negative depending on the injection angle of the quasiparicles. Loosly
speaking, the junction can be thought as a combination of so-called 0- and
$\pi$-junctions. Due to different temperature dependencies of both components
of the critical current, this may result in a nonmonotonous temperature
dependence of the Josephson current. In particular, such behaviour is expected
for symmetric [001] tilt HTS grain boundary junctions. We point out that this
feature is expected only when Andreev bound states are formed at the interface
of both junction electrodes and if there is a change in sign of the Josephson
current as a function of injection angle. The predicted nonmonotonous $I_c(T)$
is in clear contrast to the behaviour of traditional Josephson junctions
formed by $s$-wave superconductor, for which a monotonous increase of $I_c$
with decreasing temperature is theoretically predicted and experimentally
observed. The nonmonotonous $I_c(T)$ dependence for $d$-wave Josephson
junctions again is sensitive to interface roughness. For increasing roughness
the anomaly is shifted to lower temperatures \cite{Barash:96b}. This may be
the reason why this effect has not yet been convincingly experimentally
observed. Additionally, the presence of a subdominant $s$-wave component of
the order parameter as discussed in the next section is predicted to suppress
strongly the enhancement of the Josephson current at low temperatures
\cite{Tanaka:98xy}.

\subsubsection{Splitting of the zero bias conductance peak
by a broken time reversal symmetry state}

It has been shown in general that a locally time reversal symmetry breaking
state at interfaces of unconventional superconductors can exist that decays
exponentially toward the bulk \cite{Sigrist:95a}. In a broken time reversal
symmetry state the Andreev bound states shift to finite energy resulting in a
split zero bias conductance peak. A broken time reversal symmetry state is
obtained by applying a magnetic field. However, even at zero applied magnetic
field time reversal symmetry can be broken by the presence of two order
parameters at the surface having a $\pi
/2$ relative phase difference.  The possible coexistence of a second order
parameter with $s$-wave symmetry that can become stable at surfaces of HTS
but is completely dominated by the $d$-wave potential in the bulk material
has been discussed by Sigrist {\it et al.} \cite{Sigrist:95a} and Matsumoto
{\it et al.} \cite{Matsumoto:95a}. Such situation is possible at the surface
of a $d$-wave HTS, since the surface may act as strong pair breaker for the
$d$-wave paired quasiparticles thereby releasing spectral weight for the
formation of pairs in a subdominant (e.g. $s$-wave) pairing channel. Below a
surface transition temperature then a surface order parameter can develop
that spontaneously breaks time reversal symmetry and results in spontaneous
surface currents. The pairing symmetry at the surface can be a mixed $d+is$
(or in another notation $B_{1g}+iA_{2g}$) phase. Note that both order
parameters coexist at the surface with a relative phase shift of $\pi/2$ that
leads to a spontaneous surface current. The Andreev bound states then are
shifted to finite energy due to the Doppler shift ${\bf v}_f\cdot{\bf p}_s$,
where $\bf p_s$ is the condensate momentum due to the spontaneous
supercurrent. This in turn leads to a split zero bias conductance peak  in
zero applied magnetic field \cite{Fogelstroem:97a,Belzig:98}.  That is, the
measurement of the conductance in $ab$-plane tunneling experiments can give
information on the presence of a surface order parameter in HTS that
spontaneously breaks time reversal symmetry.  Recently, this effect has been
found to be consistent with the $t-J$-model \cite{Zhang:88} in the case of
sufficiently large $J$ at the surface of a $d$-wave superconductor
\cite{Tanuma:98,Kuboki:98}.

Recently, a zero field splitting and a nonlinear evolution of the magnitude
of the splitting with increasing applied magnetic field as predicted by
Fogelstr\"{o}m {\it et al.} \cite{Fogelstroem:97a}, has been reported for planar
SIN-junctions \cite{Covington:97,Lesueur:92} and for low temperature scanning
tunneling spectroscopy experiments \cite{Kashiwaya:98a}.

\subsubsection{Anomalous Meissner current}

The London-penetration depth $\lambda_L$ of a superconductor is known to be
related to the density of paired quasiparticles. For the different symmetries
of the order parameter a different behaviour of $\lambda_L(T)$ has been
predicted and experimentally found \cite{Won:94,Hardy:93,Froehlich:94}. The
presence of Andreev bound states at surfaces of HTS can have a significant
effect on the low temperature dependence of the London penetration depth. In
order to clarify this effect let us consider the surface of a $d$-wave
superconductor where Andreev bound states are formed. In a state with
established time reversal symmetry there is no net current parallel to the
surface, since in thermal equilibrium the contributions of all Andreev bound
states exactly cancel each other. However, applying a magnetic field parallel
to the surface results in a broken time reversal symmetry state and a
Meissner shielding current. This in turn results in a Doppler shift  ${\bf
v}_f \cdot {\bf p}_s$, where $\bf p_s$ now is the condensate momentum due to
the Meissner shielding current. Since the Doppler shift has opposite sign for
Andreev bound states with opposite $v_f$, in thermal equilibrium there is no
longer an equal population of states with opposite $v_f$. This results in a
net surface current that is opposite to the usual Meissner shielding current
and therefore is denoted as ``anti-Meissner'' current. In contrast to the
usual Meissner shielding current flowing in a surface layer of depth $\lambda
_L$, this ``anti-Meissner'' current flows in a surface layer given by the
$ab$-plane coherence length $\xi_{ab}$. It becomes important at low
temperatures where is causes an anomalous temperature dependence of the
London penetration depth $\lambda_L(T)$ \cite{Burkhardt:97}. Due to the
``anti-Meissner'' current the London penetration depth is slightly increased
beyond the value expected without this current. First experimental evidence
for this effect has been reported recently by Walter {\it et al.}
\cite{Walter:98} and more evidence will be presented below for HTS-GBJs.

\subsection{Blonder-Tinkham-Klapwijk model}
\label{BTK}

Blonder, Tinkham, and Klapwijk (BTK) developed a generalized Andreev
reflection model for junctions involving superconductor-normal metal (S/N)
interfaces \cite{Blonder:82,Klapwijk:82}. To account for a finite barrier
strength at the S/N interface they introduced a dimensionless factor $Z$. For
example, in the limiting case of $Z=0$ the transmission coefficient of the
junction that is given as $1/(1+Z^2)$ becomes unity. In the $Z=0$ limit and
for small values of $Z$ the BTK model predicts a peak in the normalized
differential conductance at zero voltage that reaches a maximum value of 2.
This peak decreases monotonously with increasing voltage, that is, the BTK
model predicts a conductance peak at zero bias. However, no gap structure in
the conductance versus voltage curves is expected to be observed for small
$Z$. That is, within the BTK-model the observation of both a gap like
structure {\em and} a zero bias conductance peak  is not expected.  We
emphasize that the BTK-model was developed for conventional BCS
superconductors. Therefore, in its original form the BTK-model is not
sufficient to explain the experimentally observed tunneling spectra of
$d$-wave superconductors. However, the BTK-model can be used as a starting
point for more elaborate theories. The Andreev bound state model represents a
generalized BTK-type model that accounts for an arbitrary $Z$ and an
anisotropic pair potential.

\section{Experimental Results}
\label{Exp}
\subsection{Sample preparation and experimental techniques}
\label{prep}

\begin{figure}[t]
\noindent
\vspace*{0cm}\\
\epsfig{file=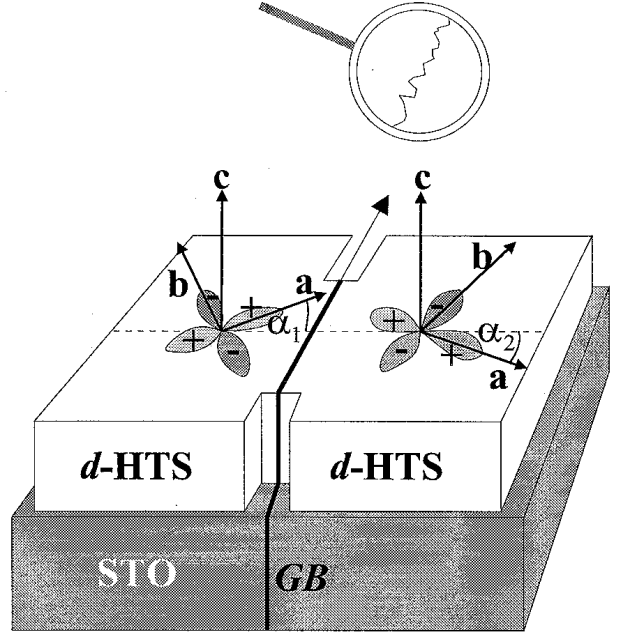,width=8.5cm,bbllx=5,bblly=760,bburx=225,bbury=1000,clip=}
\vspace*{0cm}
\caption{Sketch of the sample geometry of a symmetric [001] tilt HTS-GBJ.
The crystal axis $\vec{a}$, $\vec{b}$, and $\vec{c}$ and the misorientation
angles $\alpha _1$ and $\alpha _2$ are indicated. For a symmetric and
asymmetric junction we have $\alpha _1 = -\alpha _2$ and $|\alpha _1| \ne
|\alpha _2|$, respectively.  Also given is a magnified view of the grain
boundary microstructure indicating the strong faceting of the grain boundary.
The nominal misorientation angle is only the average angle of a wide range of
different angles of the individual facets.}
\label{gbj}
\end{figure}

In our experiments both symmetric ($\alpha _1 = - \alpha _2$) and asymmetric
($\alpha _1 =0; \;\alpha _2$) [001] tilt HTS-GBJs fabricated on SrTiO$_3$
bicrystal substrates were used. A sketch of the grain boundary junction
geometry is shown in Fig.~\ref{gbj}. The symmetric GBJs had a total
misorientation angle ($\alpha _1 + \alpha _2$) of 24$^{\circ}$ and
36.8$^{\circ}$, for the asymmetric [001] tilt GBJs we used $\alpha _2 =
45^{\circ}$. Only $c$-axis oriented epitaxial thin films were grown on the
bicrystal substrates. In this way the tunneling direction in the grain
boundary junction configurations always is along the $ab$-plane.  The YBCO
thin films were deposited using both hollow cathode magnetron sputtering and
pulsed laser deposition from stoichiometric targets. YBa$_2$Cu$_3$O$_{7-\delta
}$ samples were fabricated with different oxygen content. Both fully oxidized
YBCO with a critical temperature $T_c$ of about 90\,K and oxygen deficient
YBCO with $T_c$ values ranging between 55 and 60\,K were used. We note that
the $T_c$ reduction in the oxygen deficient YBCO is obtained by well
controlled oxygen depletion and is not caused by the presence of magnetic
impurities in the samples or at the grain boundary. Furthermore, BSCCO-GBJs
were fabricated in the same way with a $T_c$ of the BSCCO-films of about 80\,K
\cite{Mayer:93}. This $T_c$ value is reduced by about 10\ K as compared to
optimum doped samples \cite{Marx:95}. The epitaxial LSCO thin films have been
grown on the bicrystal substrates by reactive coevaporation from metal sources
using ozone as reaction gas \cite{Naito:95,Beck:96a,Sato:97}. The $T_c$ of
these samples typically was about 24\,K. This value is comparable to other
epitaxial thin film samples but slightly reduced compared to that of the best
single crystals of LSCO. This most likely is due to strain at the interface
between the substrate and the epitaxial thin film \cite{Sato:97,Locquet:96}.
Finally, the NCCO films were deposited similar to the LSCO films by molecular
beam epitaxy (MBE) using ozone as oxidation gas \cite{Naito:95}. The $T_c$ of
about 24\ K is one of the highest reported for this material in literature
including even NCCO single crystals \cite{Yamamoto:97}. After their growth,
the HTS thin film bicrystals have been patterned using optical lithography and
Ar ion beam etching to obtain a GBJ structure as sketched in Fig.~\ref{gbj}.

The measurements of the current-voltage [$I(V)$] and conductance vs.~voltage
[$G(V)=dI(V)/dV$] characteristics were performed in a standard four-probe
arrangement. If necessary the critical current was suppressed by applying a
small magnetic field parallel to the grain boundary plane corresponding to a
minimum in $I_c$ vs.~$H$. Zero field measurements were performed in
magnetically shielded cryostats placed in an rf-shielded room. Also, high
magnetic fields up to 12\,T and low temperatures down to 100\,mK were used in
our study.

%%%%%%%%%%%%%%%%%%%%%%%%  FIGURE 3 %%%%%%%%%%%%%%%%%%%%%%%%%

%%%%%%%%%%%%%%%%%%%%%%%%%%%%%%%%%%%%%%%%%%%%%%%%%%%%%%%%%%%%

\subsection{Transport mechanism in GBJs}
\label{gbjs}

The electrical transport properties of grain boundaries in the cuprate
superconductors have been studied intensively over the last years. Recent
reviews have been given by Gross {\em et al.} \cite{Gross:94a,Gross:97a}.
Here, we briefly summarize the main results relevant to the tunneling
spectroscopy experiments discussed below. Phenomenologically, the
superconducting properties of the GBJs can be well modeled by the resistively
and capacitively shunted junction (RCSJ) model \cite{Gross:94a,Likharev:86}.
In order to describe the vaste majority of the present experimental data on
the electrical transport and noise properites of HTS-GBJs the intrinsically
shunted junction (ISJ) model has been suggested \cite{Gross:91m,Gross:92book}.
In this model it is assumed that there is a thin insulating grain boundary
barrier containing a large density of localized defect states. The microscopic
transport mechanism then is direct and resonant tunneling via the localized
states. A key feature of the ISJ-model is the fact that the transport of
Cooper pairs via the localized states is prevented by Coulomb repulsion, i.e.
Cooper pairs have to use the direct tunneling channel, whereas due to the
large density of defect states the quasiparticle transport is dominated by
resonant tunneling via a single localized state
\cite{Marx:95,Halbritter:93,Froehlich:95,Alff:96pc,Froehlich:97a,Marx:97,Kleefisch:98}.
The resonant channel can be viewed as to provide an intrinsic resistive shunt
leading to the name intrinsically shunted junction. The ISJ-model well
explains the scaling behaviour $V_c=J_c\rho _n \propto (J_c)^q$ with $q \sim
0.5$ observed for the GBJs \cite{Gross:91m,Gross:90s}. Here, $J_c$ and $\rho
_n$ are the critical current density and the normal resistance times area of
the GBJ. Furthermore, it naturally accounts for the large amount of low
frequency excess noise due to trapping and release of charge carriers within
the localized states \cite{Marx:95,Marx:97}.

We emphasize that the tunneling processes present in GBJs are elastic
processes. If inelastic transport processes would be important, a significant
temperature dependence of the tunneling conductance for voltages well above
the gap voltage would be expected. However, the experimental data on GBJs show
a temperature independent conductance \cite{Froehlich:97}; i.~e.~inelastic
tunneling via two and more localized centers is negligible. Hence, the
tunneling quasiparticles can carry spectroscopic information on the density of
states (DOS) in the junction electrodes and thereby e.g. on the energy gap. In
contrast, such spectroscopic information is destroyed, if multistep inelastic
tunneling or even variable range hopping would be the dominating transport
process. We also note that the localized states within the grain boundary
barrier have an about white energy distribution. Hence, they do not act as an
energy filter producing structures in the conductance versus voltage curves of
HTS-GBJs. Because all junctions used in our experiments could be well fitted
within the above picture, we assume that the same tunneling mechanism is
present in all investigated junctions.

%%%%%%%%%%%%%%%%%%%%%%%%% FIGURE 4 %%%%%%%%%%%%%%%%%%%%%%%%%

\begin{figure}[hbt]
\noindent
\vspace*{-0.5cm}\\
\epsfig{file=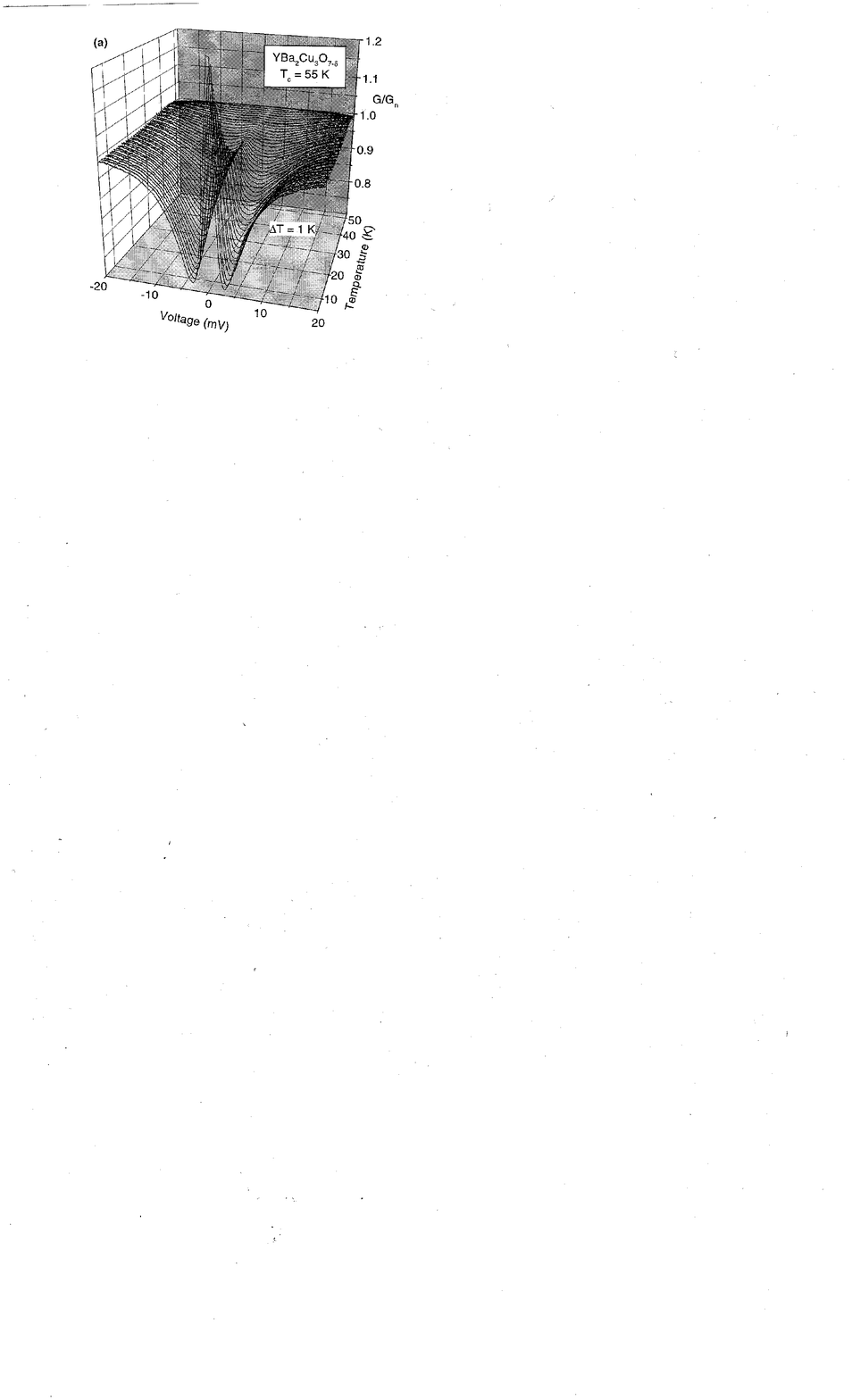,width=8.5cm,bbllx=55,bblly=765,bburx=290,bbury=1020,clip=}
\hspace*{0cm}\centering\epsfxsize=12.5cm\epsffile{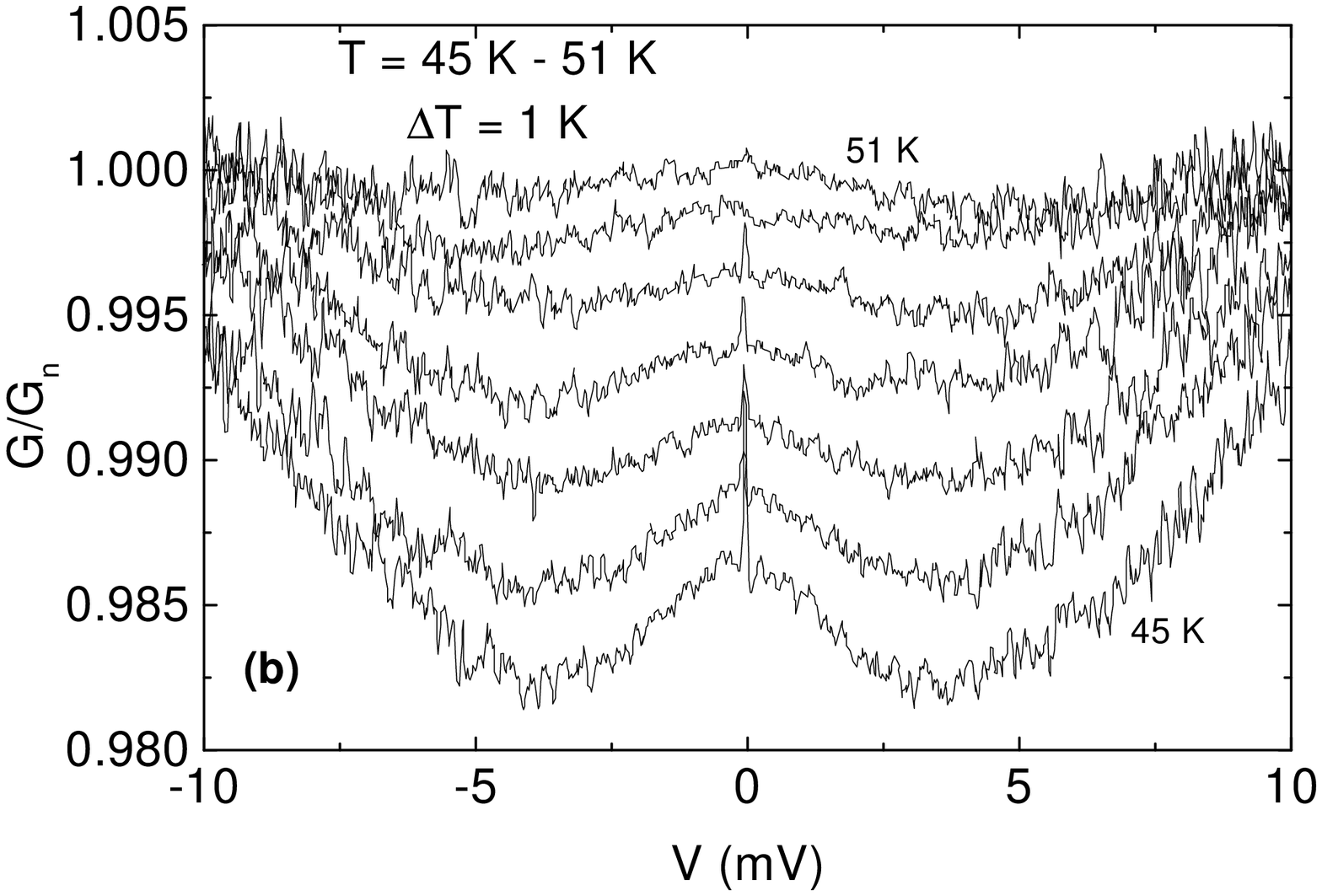}
\vspace*{-3.5cm}
\caption{{\bf (a)} Normalized differential conductance vs.~voltage for a $24^{\circ}$ [001] tilt
YBCO-GBJ for temperatures between 4 and 51\ K. The sample was underdoped with
$T_c\approx 55$\ K. Some data points close to zero are removed because of the
Josephson current peak. {\bf (b)} The same dependence for temperatures
between 45 and 51\ K on an enlarged scale. One clearly can distinguish the
sharp Josephson current peak at zero bias, the much broader zero bias
conductance peak, and the superconducting gap structure even for $T$ close to
$T_c$.}
\label{epj98f3}
\end{figure}

%%%%%%%%%%%%%%%%%%%%%%%%%%%%%%%%%%%%%%%%%%%%%%%%%%%%%%%%%%%%

There have been extensive studies on the microstructure of the grain boundary
barrier. High resolution transmission electron microscopy (HRTEM) revealed a
strongly faceted grain boundary barrier with facets on a nanoscale
\cite{Kabius:94}. This means that the macroscopic misorientation angles
$\alpha _1$ and $\alpha _2$ of the grain boundary, which are determined by
the bicrystal substrate, represent average values only. However, on a
microscopic scale these angles can vary considerably due to the faceting with
$|\alpha _1| + |\alpha _2| = const$.  As a consequence in some cases GBJs
formed by $d$-wave HTS materials have to be modelled as a parallel array of
so-called ``0''- and ``$\pi$''-junctions showing an irregular magnetic field
dependence of the critical current \cite{Mannhart:96,Hilgenkamp:96}. This
effect is negligible for symmetric GBJs with small misorientation angles
($|\alpha_1|+|\alpha_2|<25^{\circ}$) \cite{Kleefisch:98}, however, it becomes
more pronounced for increasing misorientation angle and stronger faceting. In
general, the faceted grain boundary barrier can be viewed as a rough
interface. The impact of surface roughness on the spectral weight of Andreev
bound states on their consequences already has been discussed above.
Unfortunately, up to now no exact quantitative description of the effects of
a faceted grain boundary on the electrical transport properties is available.
Below we will qualitatively discuss possible effects of the faceting that can
be understood from simple angle averaging.

\subsection{Tunneling spectroscopy of HTS-GBJs}
\label{tunnel}

In this section we present our experimental results on the tunneling
spectroscopy on various HTS materials using GBJs. In the next section we will
discuss in detail our results in the framework  of the different models
discussed above.

We first consider the conductance versus voltage curves of the hole doped HTS
YBCO, BSCCO, and LSCO, which are believed to have a $d$-wave symmetry of the
order parameter. Fig.~\ref{epj98f3}a shows a set of typical $G(V)$-curves
normalized to the normal state conductance $G_n$. The curves were obtained
for a 24$^{\circ}$ [001] tilt YBCO-GBJ at temperatures ranging between 4 and
51\,K. The Josephson current peak at zero voltage has been removed for
clarity. The sample of Fig.~\ref{epj98f3}a was underdoped and had a $T_c$ of
about 55\,K. In Fig.~\ref{epj98f3}b some spectra are shown on an enlarged
scale in order to better show the presence of the broad zero bias conductance
peak  up to the $T_c$ of the material. Furthermore, the superconducting gap
structure and a narrow peak at $V=0$ due to the Josephson current can be
observed up to $T_c$. For other hole doped HTS materials such as BSCCO and
LSCO the same behaviour was observed. For LSCO with $T_c\approx24$\,K the
experimental results have been published recently \cite{Alff:98x}. For all
samples the zero bias conductance peak  is found to decrease with increasing
temperature and to disappear at $T_c$. Furthermore, also the gap structure is
found to disappear at $T_c$.

The measured tunneling spectra are almost perfectly symmetric about zero
voltage as expected for a symmetrical tunnel junction with electrodes formed
by the same material. Far above the gap voltage the conductance was found to
be independent on temperature and has an about parabolic shape
\cite{Froehlich:97a}. The parabolic shape can be explained by the effect of
the applied voltage on the shape of the barrier potential. The background
conductance can be viewed as normal state conductance $G_n$. In order to show
the temperature dependence of the spectra more clearly, the conductance data
have been normalized with respect to $G_n$. Fig.~\ref{epj98f3} clearly shows
that below the gap voltage the density of states is reduced by about 30\%.
With increasing temperature the conductance smoothly merges towards to the
normal state conductance when the temperature is approaching $T_c$. Recently,
Ekin {\it et al.} \cite{Ekin:97} reported the disappearance of the zero bias
conductance peak  at temperatures well below the transition temperature of
the material. Such behaviour has not been observed in our study and thus
might be an effect of sample fabrication. In our case a clear correlation
between the appearance of a measurable zero bias conductance peak  and $T_c$
was observed.

\begin{figure}[t]
\noindent
\vspace*{0cm}\\
\hspace*{0cm}\centering\epsfxsize=12.5cm\epsffile{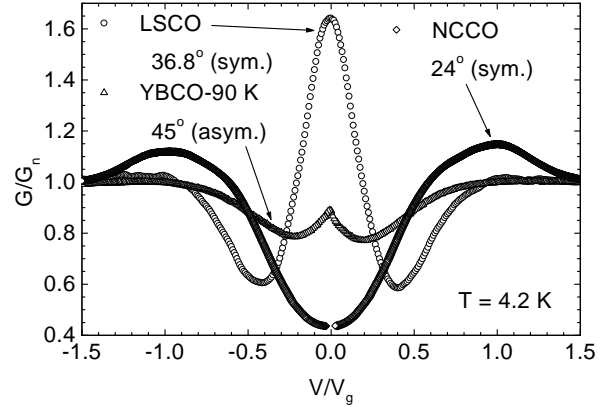}
\vspace*{-3.5cm}
\caption{Normalized conductance vs.~normalized voltage of GBJs formed by
LSCO, YBCO (90\,K phase), and NCCO. The YBCO-GBJ was an asymmetric [001] tilt
GBJ while the LSCO- and NCCO-GBJ were symmetric [001] tilt GBJs. For the
NCCO-GBJ the Josephson current peak has been removed.}
\label{epj98f4}
\end{figure}

The materials we used for our experiments were both hole doped HTS (60 and
90\,K phase of YBCO, BSCCO, and LSCO) and an electron doped HTS (NCCO). We now
compare the data of the hole doped HTS to that obtained for the electron doped
HTS. There is significant experimental evidence that the hole doped materials
have a $d$-wave symmetry of the order parameter, whereas NCCO has a dominating
$s$-wave component. From our experiments on GBJs fabricated from these
materials we can conclude that all hole doped HTS show a pronounced zero bias
conductance peak, whereas such peak is completely absent for the electron
doped NCCO.  In order to demonstrate this observation, in Fig.~\ref{epj98f4}
we show data for fully oxidized YBCO, as well as for LSCO and NCCO. Again the
data is normalized to the normal conductance $G_n$. Furthermore, the voltage
scale is normalized to the gap voltage $V_g$ in order to compare data of
materials with different gap values. Here, we have chosen $V_g = \Delta_0/e$
and not, as for tunnel junction using conventional BCS-like superconductors,
$V_g=2\Delta_0/e$. We also note that the exact position of the conductance
peak due to the gap depends on the orientation of the electrodes with respect
to the barrier. This is a consequence of the (interface) density of states in
a $d$-wave superconductor \cite{Barash:95a}. In general, the same
consideration should apply for a highly anisotropic $s$-wave superconductor.
Therefore, in a first order approximation $V_g \sim
\Delta_0/e$ is assumed for NCCO as well. Typical gap voltages of the
different materials were 6\,mV for both LSCO and NCCO, 15\,mV for the 60\,K
phase of YBCO (YBCO-60), 20\,mV for the 90\,K phase (YBCO-90), and 25\,mV for
BSCCO.

%%%%%%%%%%%%%%%%%%%%%%%%  FIGURE 5 %%%%%%%%%%%%%%%%%%%%%%%%%

%%%%%%%%%%%%%%%%%%%%%%%%%%%%%%%%%%%%%%%%%%%%%%%%%%%%%%%%%%%%

With respect to the gap structure in the measured tunneling spectra we would
like to add an interesting observation concerning the magnitude of the gap. As
can be seen from Fig.~\ref{epj98f3}a the magnitude of the gap seems to be
about constant with increasing temperature. However, the states within the gap
are filled up with increasing temperature until the normal state value is
reached. Above $T_c$ the presence of a gap could {\em not} be observed in our
experiments. A similar behaviour has been observed by Renner {\it al.}
\cite{Renner:96,Renner:98} using low temperature scanning tunneling
spectroscopy. However, in contrast to our observation these authors report a
gap persisting even above $T_c$. We note, that the tunneling direction in the
experiments by Renner {\it et al.} was along the $c$-axis, whereas it is along
the $ab$-plane in our study.  The observed temperature evolution of the gap
structure completely deviates from the BCS-behaviour and is probably related
to the pseudogap that has been observed in the HTS \cite{Ding:96,Loeser:96}.

We next consider the magnitude of the gap as a function of the doping level.
Recently, an increasing gap was found with decreasing doping concentration for
BSCCO \cite{Renner:98,Miyakawa:98}. In the case of YBa$_2$Cu$_3$O$_{7-\delta
}$ we investigated the two doping concentrations $\delta \simeq 0$ and $\delta
\simeq 0.4$. In contrast to the recent work on BSCCO, we observed a linear scaling of
the gap values with $T_c$. We note, however, that this behaviour has to be
examined in more detail in future. Also for NCCO, the gap value was found to
scale with $T_c$.  A junction with a $T_c$ of about 14\,K had a gap value of
about $3.5$\,meV, whereas  junctions with $T_c\simeq 24$\,K showed gap values
of about 6\,meV.

The conductance versus voltage curves of GBJs fabricated from the $d$-wave HTS
\cite{Achsaf:96} all show some common properties. Firstly, they have a
pronounced zero bias conductance peak. Secondly, the width of the zero bias
conductance peak  is always about several mV and is not narrowing with
decreasing temperature. Thirdly, the conductance is reduced in an energy range
corresponding to the superconducting gap energy most likely due to a reduced
density of states. For all $d$-wave HTS only a reduction of the density of
states of about 30 to 50\% is observed. Additionally, a broad and flat peak in
the $G(V)$ curves is observed at the gap voltage, however, not for all
samples. Finally, reducing the temperature to very small values (down to
100\,mK) results in a zero bias conductance peak  with a shape that is more
Lorentzian-like than Gaussian-like. As shown in Fig.~\ref{epj98f4}, a Gaussian
shape is obtained for LSCO at 4.2\,K, whereas for YBCO-90 already a Lorentzian
shape is obtained at this temperature. This indicates that the measured
linewidth is not determined by extrinsic effects such as thermal smearing or
the finite experimental resolution but represents an intrinsic linewidth of
the zero bias conductance peak. Calculating the linewidth of the zero bias
conductance peak, an imaginary part can be added to the energy introducing a
phenomenological lifetime parameter $\Gamma$ of the order of $\Delta_0/10$
\cite{Alff:97b}. Recently, a momentum-dependent intrinsic broadening of the
surface bound states has been predicted \cite{Walker:98}.

The behaviour of NCCO-GBJs differs from that of GBJs fabricated from the
$d$-wave HTS in several features. Firstly, the $G(V)$ curves of optimum doped
NCCO-GBJs with a maximum $T_c$ of 24\,K show a much clearer peak structure at
$V_g$ than those of the $d$-wave HTS-GBJs. Secondly, the gap structure is
more pronounced and the density of states is reduced to about 40\% of the
normal state value. For a NCCO-GBJ with a reduced $T_c$ both the peak at
$V_g$ and the reduction of the density of states becomes smaller. However, up
to now no detailed tunneling spectroscopy as a function of doping has been
performed. Finally, it is interesting to note that the spectra of NCCO have,
apart from the zero bias conductance peak, more similarities to those of the
hole doped HTS materials than to those of conventional BCS-superconductors.

%%%%%%%%%%%%%%%%%%%%%%%%  FIGURE 6 %%%%%%%%%%%%%%%%%%%%%%%%%

\begin{figure}[htb]
\noindent
\vspace*{0cm}\\
\hspace*{0cm}\centering\epsfxsize=12.5cm\epsffile{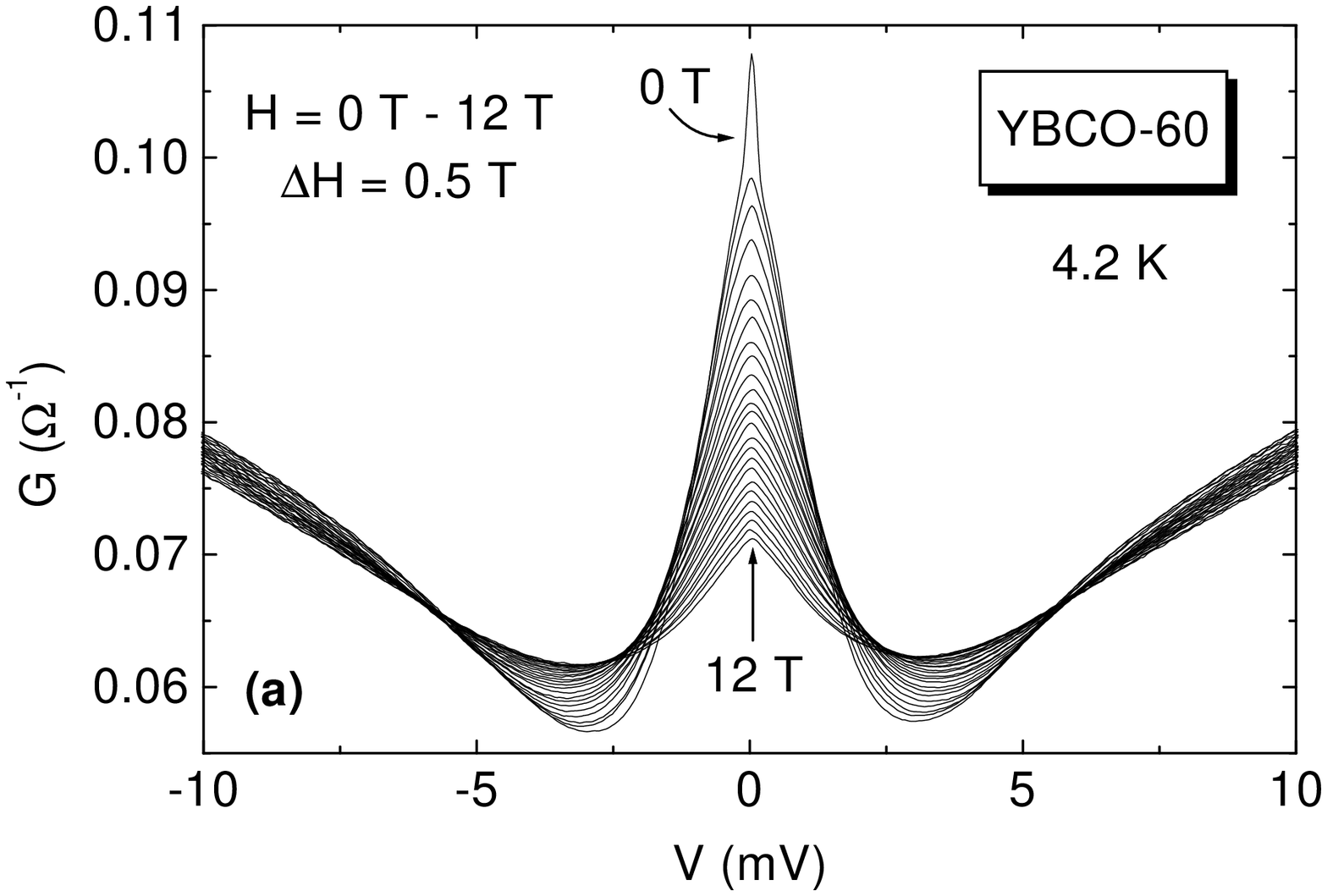}\\[-3cm]
\hspace*{0cm}\centering\epsfxsize=12.5cm\epsffile{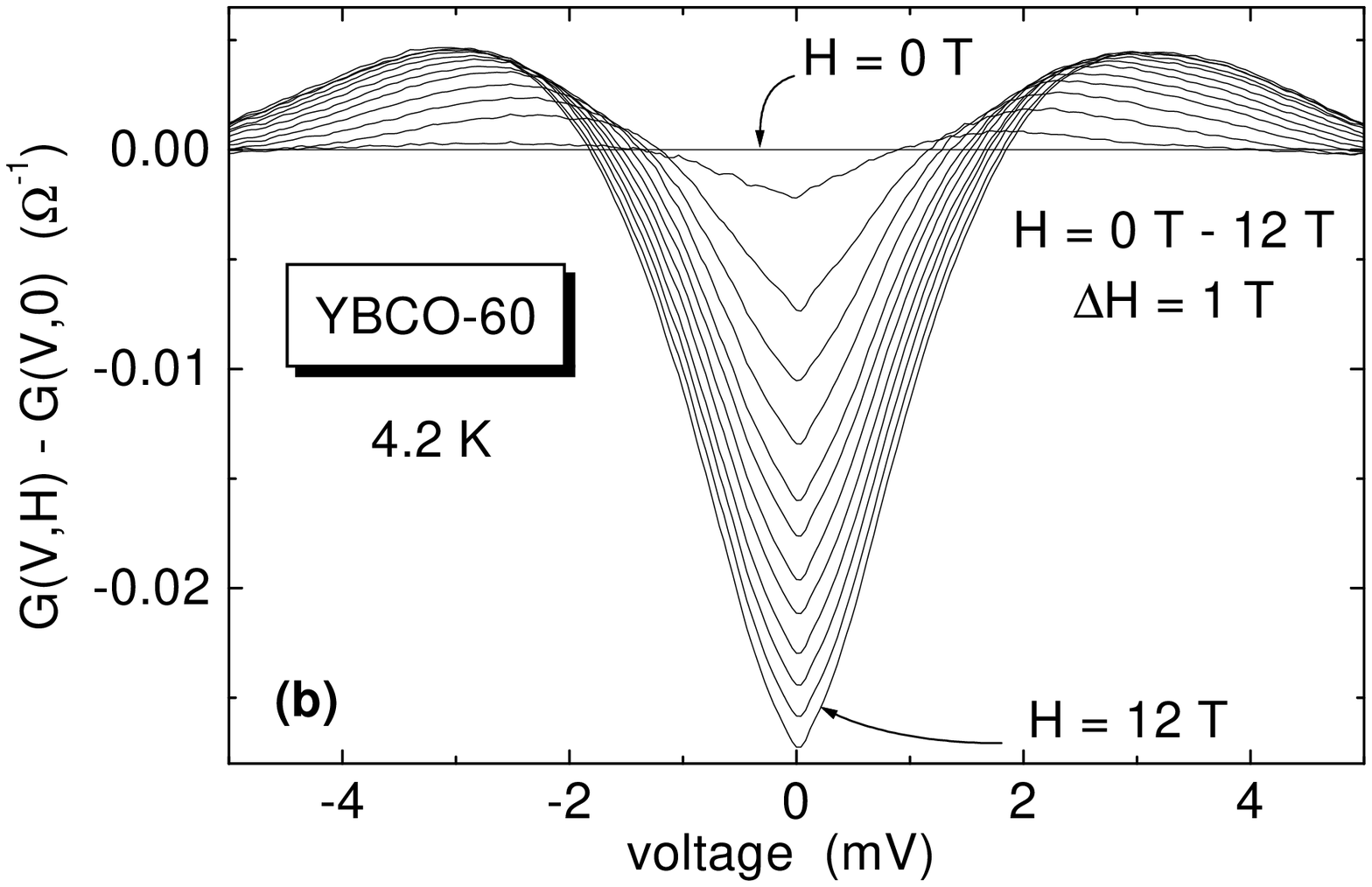}
\vspace*{-3.5cm}
\caption{(a) Magnetic field dependence of the zero bias conductance peak  at 4.2\,K for a symmetric
[001] tilt YBCO-GBJ (60\ K phase). In (b) $(G(V,H)-G(V,0)$ is plotted for the
same sample. The applied magnetic fields ranged between 0 and 12\,T ($0.5$\,T
steps).}
\label{epj98f5}
\end{figure}

%%%%%%%%%%%%%%%%%%%%%%%%%%%%%%%%%%%%%%%%%%%%%%%%%%%%%%%%%%%%

We finally consider the magnetic field dependence of the conductance versus
voltage curves.  The magnetic field dependence of the tunneling spectra of a
YBCO-GBJ (60\ K phase) is shown in Fig.~\ref{epj98f5}(a). The magnetic field
$H$ was applied parallel to the grain boundary barrier in $c$-axis direction.
The magnetic field was increased in 0.5\,T steps. In Fig.~\ref{epj98f5}(a)
the Josephson current has not been suppressed explaining the sharp spike at
zero voltage in the zero field $G(V)$ curve on top of the much broader zero
bias conductance peak caused by the Andreev bound states. The effect of the
applied magnetic field is the suppression of spectral weight around zero
energy. The spectral weight removed around zero energy is shifted to higher
energies about the whole energy range. Hence, in the $G(V,H)-G(V,0)$
vs.~voltage curves shown in Fig.~\ref{epj98f5}(b) two peaks appear with the
peak position increasing non-linearly with increasing applied magnetic field
\cite{Froehlich:97a}. We note that the overall spectral weight at 12\,T is
reduced by about 3\% below the value at zero magnetic field if we consider a
voltage interval of $\pm$12\,mV. However, it is not likely that the
conservation of the number of states is violated but that the missing
spectral weight is shifted to even higher energies not contained in the
considered voltage interval. For LSCO at 100\ mK qualitatively the same
result has been obtained \cite{Alff:98x}. Note that the thermal smearing is
only of the order of 10\ $\mu$eV at 100\ mK. This explains that for LSCO the
zero bias conductance peak  in the conductance vs. voltage curves changes
from a Gaussian shape at 4.2\ K to Lorentzian shape at 100\,mK. In contrast
to reference \cite{Covington:97} but in accordance to reference
\cite{Ekin:97} no direct splitting of the zero bias conductance peak  was
found at any applied magnetic field up to 12\,T.

\subsection{Comparison to low temperature scanning tunneling spectroscopy measurements}
\label{LTSTS}

\begin{figure}[t]
\noindent
\vspace*{0cm}\\
\hspace*{0cm}\centering\epsfxsize=12.5cm\epsffile{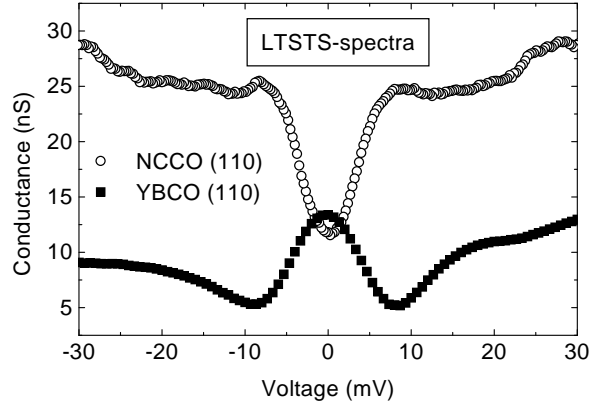}
\vspace*{-3.5cm}
\caption{Low temperature scanning tunneling spectroscopy spectra for (110) oriented surfaces of NCCO and YBCO at 4.2\,K.
The data are taken from reference \cite{Alff:97c}.}
\label{epj98f6}
\end{figure}

In this subsection we briefly compare the results obtained for SIS-type
HTS-GBJs to those obtained using low temperature scanning tunneling
spectroscopy, where a SIN-configuration is used. A low temperature scanning
tunneling spectroscopy study of different HTS materials with different
orientations has been given recently by Alff {\it et al.}
\cite{Alff:96b,Alff:97c}. Details on the low temperature scanning tunneling
spectroscopy technique can be found in \cite{Alff:96b,Alff:97c} and in the
references cited therein. Typical conductance versus voltage curves obtained
by low temperature scanning tunneling spectroscopy are shown in
Fig.~\ref{epj98f6}. These curves can directly be compared to those of
Fig.~~\ref{epj98f4} obtained from experiments using GBJs.  We note that the
conductance of the two junction types differs by several orders of magnitude.
Whereas one has a typical conductance of $10^{-8}$\,S in low temperature
scanning tunneling spectroscopy spectra, the conductance of GBJs ranges
between $10^{-1}$ and $10^{-2}$\,S. However, this difference mainly is caused
by the different junction area, which is by several orders of magnitude
smaller for low temperature scanning tunneling spectroscopy. The low
temperature scanning tunneling spectroscopy data of Fig.~\ref{epj98f6} were
obtained by performing experiments on (110) oriented surfaces of NCCO and
YBCO. With respect to the zero bias conductance peak, the same result is
obtained as for the GBJs. Again, only YBCO shows a zero bias conductance peak
while such peak is completely absent for NCCO. We note that by low temperature
scanning tunneling spectroscopy also (100) and (001) oriented surfaces have
been probed. Also for these surfaces a zero bias conductance peak  never has
been observed for NCCO \cite{Alff:96b}. However, for YBCO in many cases a zero
bias conductance peak  could be detected for (100) and (001) oriented
surfaces. For a (100) oriented surface, where in the ideal case no zero bias
conductance peak  is expected, this is most likely related to the finite
roughness of the surface as discussed in section~\ref{Observe}. This effect is
similar to the observation of zero bias conductance peaks for $a$-axis or even
$c$-axis oriented planar junctions.

%%%%%%%%%%%%%%%%%%%%%%%%  FIGURE 7 %%%%%%%%%%%%%%%%%%%%%%%%%

%%%%%%%%%%%%%%%%%%%%%%%%%%%%%%%%%%%%%%%%%%%%%%%%%%%%%%%%%%%%

We also would like to address another detail of the conductance versus
voltage curves that are similar for both the SIN junctions used in low
temperature scanning tunneling spectroscopy and the SIS-type GBJs. For NCCO
the $G(V)$ curves show a more pronounced superconducting gap structure with a
density of states that is reduced by about 50\% within the gap. In contrast,
for YBCO this reduction is smaller and (if at all) at the gap voltage only
very faint and broad conductance peaks are observed. Finally, we note that in
contrast to the spectra of the GBJs the low temperature scanning tunneling
spectroscopy spectra show some asymmetry that is most pronounced in the case
of YBCO. This feature has been observed in many spectra using low temperature
scanning tunneling spectroscopy \cite{Renner:95,Nantoh:95}. So far, it is not
clear whether this effect is related to the order parameter symmetry.
However, one is led to think in this direction because the asymmetry is
absent for NCCO.

\subsection{Anomalous Meissner currents}
\label{amc}

%%%%%%%%%%%%%%%%%%%%%%%%  FIGURE 8 %%%%%%%%%%%%%%%%%%%%%%%%%

\begin{figure}[b]
\noindent
\vspace*{0cm}\\
\hspace*{0cm}\centering\epsfxsize=12.5cm\epsffile{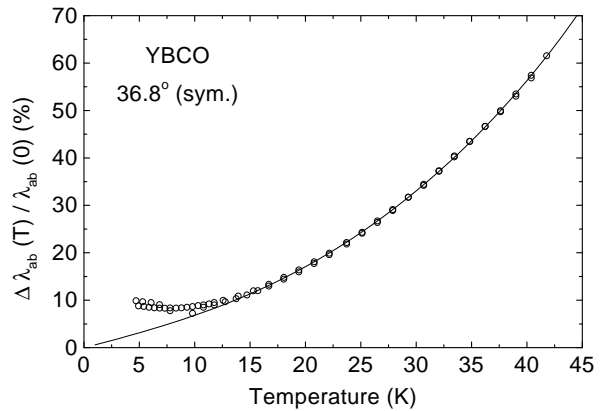}
\vspace*{-3.5cm}
\caption{$[\lambda_{ab}(T)-\lambda_{ab}(0)]/\lambda_{ab}(0)$ for a 36.8$^{\circ}$
[001] tilt symmetric YBCO-GBJ. The solid line shows the dependence expected
for a superconductor with a $d_{x^2-y2}$-symmetry of the order parameter. The
measured upturn at low temperatures most likely is due to anomalous Meissner
currents.}
\label{epj98f75}
\end{figure}

%%%%%%%%%%%%%%%%%%%%%%%%%%%%%%%%%%%%%%%%%%%%%%%%%%%%%%%%%%%%

GBJs formed by different HTS have been successfully used to determine the
relative change
\[
\displaystyle\frac{\Delta\lambda_{ab}(T)}{\lambda_{ab}(0)} =
\displaystyle\frac{\lambda_{ab}(T) - \lambda_{ab}(0)}{\lambda_{ab}(0)}
\]
of the in-plane London penetration depth $\lambda_{ab}$ with the high
precision of only a few \AA \cite{Froehlich:94}. In this technique the shift
of the side-lobes of the $I_c(H)$ dependence of HTS-GBJs is measured as a
function of temperature. Details of this precision measurement technique have
been described elsewhere \cite{Froehlich:94}. Recently, an anomalous
temperature dependence of $\Delta\lambda_{ab}$ was measured for a symmetric
36$^{\circ}$ [001] tilt YBCO-GBJ \cite{Zittartz:97}. As can be seen in
Fig.~\ref{epj98f75}, for temperatures below about 10\,K the
$\Delta\lambda_{ab}(T)/\lambda_{ab}(0)$ dependence shows a clear upturn with
decreasing temperature in contradiction to the behaviour expected for a
$d$-wave superconductor. We note, that this anomaly can be explained neither
by assuming a pure $d$-wave nor a pure $s$-wave symmetry of the order
parameter. In both cases $\Delta\lambda_{ab}(T)/\lambda_{ab}(0)$ is expected
to monotonously decrease with decreasing temperature. However, as discussed in
section~\ref{Observe} Andreev bound states can result in an anomalous Meissner
current that is flowing in opposite of the usual Meissner current thereby
increasing the measured penetration depth. This in turn results in an upturn
of the temperature dependence of the London penetration depth at low
temperatures. We emphasize, however, that a clear upturn  as shown in
Fig.~\ref{epj98f75} has not been observed for all investigated GBJs. At
present it is not known what determines the strength of the observed upturn.
However, we suppose that interface roughness is an important factor.
Additional experiments are required to clarify this issue.

\section{Discussion}
\label{Discussion}
\subsection{Magnetic impurity scenario}
\label{MIS}

Initially, the Anderson-Appelbaum model \cite{Appelbaum:66,Anderson:66} was
used by most groups to give an interpretation of the observed zero bias
conductance peak  in tunneling spectroscopy experiments on HTS
\cite{Covington:96,Walsh:92,Lesueur:92,Kashiwaya:94,Froehlich:97}. However,
later on it became clear that the Appelbaum-Anderson model can explain only a
few features of the experimental observations. For example, the predicted
logarithmic dependence of the conductance on voltage as given by
Eq.~(\ref{aa}) has been observed for example in Ref.~\cite{Froehlich:97}. We
note, however, that according to Eq.~(\ref{aa}), thermal smearing prevents the
observation of a logarithmic dependence for $eV \le k_BT$, i.e. for $V$
smaller than about 1\,mV at a measuring temperature of a few K. Therefore, the
logarithmic dependence was verified only for a relatively small voltage/energy
scale. Hence, even though the measured data are in fair agreement with the
Appelbaum-Anderson prediction, this cannot be considered a definite proof for
the validity of this model for the description of the observed zero bias
conductance peak.

We now briefly summarize the main arguments against the validity of the
Appelbaum-Anderson model. Firstly, a strong argument against the applicability
of the Appel\-baum-Anderson model is the clear correlation of the temperature
for which the zero bias conductance peak  disappears and the critical
temperature of the investigated HTS that varies between 20 and 90\,K. Within
the Appelbaum-Anderson model, no such correlation is expected, since the
quasiparticle scattering off magnetic impurities leading to the zero bias
conductance peak  within this model is independent of the onset of
superconductivity. In the Appelbaum-Anderson model there is no distinct
temperature where the zero bias conductance peak  is supposed to disappear.
Secondly, the Appelbaum-Anderson model cannot account for the fact that for
NCCO a zero bias conductance peak  never could be observed. If Cu$^{2+}$-ions
at the grain boundary interface are supposed to act as magnetic impurities
causing the zero bias conductance peak, then it is very difficult to explain
that the zero bias conductance peak appears only for YBCO, BSCCO, and LSCO,
whereas it is absent for NCCO that contains the same copper-oxygen layers as
the basic structural element responsible for superconductivity. We also note
that the presence of magnetic scatterers at the grain boundary interface is
not related to the presence of a superconducting region with a magnetic
impurity concentration $n_m$ \cite{Ovchinnikov:96}. This would include a
significant reduction of the critical temperature $T_c$ that is clearly not
observed for all of the GBJs.

We next discuss the behaviour in the presence of an applied magnetic field.
Provided that there are magnetic impurities at the grain boundary, an applied
magnetic field is expected to result in a Zeeman splitting of the impurity
levels of $2\delta=2g\mu_BH$. This has been observed for example in the case
of Ta-TaO$_2$-Al junctions \cite{Appelbaum:72}. In our GBJ experiments an
indirect splitting of the zero bias conductance peak  in an applied magnetic
field has been observed. Here, indirect means that a splitting could be seen
only in the $G(V,H) - G(V,0)$ curve, i.e. after subtracting the zero field
curve. However, firstly at low fields this splitting is much too large than
expected according to the Appelbaum-Anderson model and secondly, at larger
fields a nonlinear increase of the splitting with increasing applied field is
observed. In order to model this behaviour within the Appelbaum-Anderson model
a strongly magnetic field dependent $g$-factor would be required that at low
fields is an order of magnitude larger than that observed for metal - metal
oxide - metal junctions. For the latter, typically $g\sim 1-3$ independent of
the applied magnetic field is obtained.  Moreover, in Ref.~\cite{Covington:97}
a finite splitting of the zero bias conductance peak was observed for a
SIN-junction already in zero field. Summarizing our discussion we clearly can
state that the behaviour of the zero bias conductance peak  in the presence of
an applied magnetic field cannot be explained within the Appelbaum-Anderson
model.

In summary, for all spectroscopic experiments on HTS junctions reporting the
observation of zero bias conductance peaks there are strong arguments against
the applicability of the Appelbaum-Anderson model \cite{Hu:98}. This statement
holds for all investigated junction types with at least one HTS electrode,
i.e. HTS/I/N-type junctions used in low temperature scanning tunneling
spectroscopy, planar HTS/I/N junctions, and HTS/I/HTS type GBJs.

\subsection{BTK-model}
\label{BTKa}

As discused in section~\ref{BTK}, for small values of $Z$ the BTK model
predicts a peak in the normalized differential conductance at zero voltage
that reaches a maximum value of 2. However, this peak should decrease
monotonously with increasing voltage and no gap structure is expected to be
observed. That is, within the BTK-model the observation of both a gap like
structure {\em and} a zero bias conductance peak  is not expected. However,
in all the GBJ experiments a clear gap structure and a pronounced zero bias
conductance peak  always is observed at the same time. Furthermore, in the
GBJs no subharmonic gap structures at voltages $2\Delta/n$ were observed that
are predicted by the BTK-model. Finally, the BTK-model does not account for
the clear differences between the $s$-wave material NCCO and the $d$-wave
materials YBCO, BSCCO, and LSCO. This is evident because the BTK-model was
developed for conventional BCS superconductors and therefore cannot account
for effects related to $d$-wave superconductivity. It has been pointed out
above that the Andreev bound state model represents a generalized BTK-type
model that accounts for an arbitrary $Z$ and an anisotropic pair potential.
As discussed in the next section this model well describes the experimental
data.

\subsection{Andreev bound states in $d$-wave superconductors}
\label{ABSd}

In the following we interpret the tunneling experiments on HTS within the
Andreev bound state model. We will see that the two essential experimental
findings for the HTS-GBJs are naturally explained by this model. Firstly, the
clear correlation of the appearance of the zero bias conductance peak  in the
$G(V)$ curves with the critical temperature is a strong hint that the presence
of the zero bias conductance peak  is related to the presence of
superconductivity. Secondly, the observation that only for $d$-wave
superconductors the zero bias conductance peak  is observed gives strong
evidence for the correlation of this effect with the symmetry of the order
parameter. For NCCO, that is believed to be an $s$-wave superconductor, and
also for conventional low temperature superconductors a zero bias conductance
peak  could not be observed. This is fully consistent with the Andreev bound
state model.  In addition, the Andreev bound state model provides a convincing
way to explain almost all experimental observations of zero bias conductance
peaks in different experimental situations within a {\em single} model
\cite{Hu:98}.

\begin{figure}[htb]
\noindent
\vspace*{0cm}\\
\hspace{0cm}\centering\epsfxsize=12.5cm\epsffile{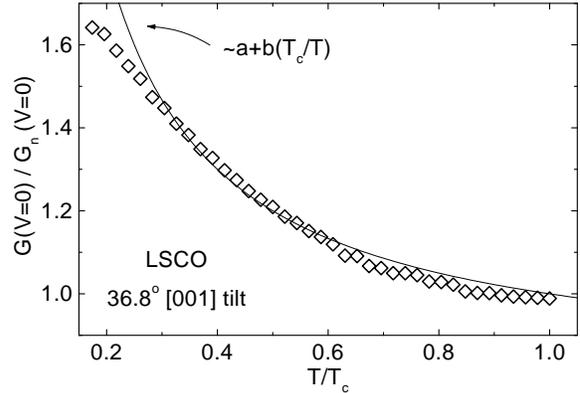}
\vspace*{-3.5cm}
\caption{Normalized conductance at zero bias of a symmetric
36.8$^{o}$ [001] tilt LSCO-GBJ plotted vs.~normalized temperature.}
\label{epj98f7}
\end{figure}

We first discuss the temperature dependence of the zero bias conductance
peak. However, we note that for this dependence so far no complete
theoretical description is available at present. In Fig.~\ref{epj98f7} the
normalized conductance at zero bias, $G(0)/G_n$,  of a LSCO-GBJ is plotted
versus the reduced temperature $t=T/T_c$. For $t>0.3$, the functional form is
close to a $1/T$ dependence. Recently, such dependence has been predicted by
Barash {\it et al.} \cite{Barash:97}. However, these authors predict a $1/T$
dependence for small values of $t$, whereas we find such dependence for large
$t$ and, moreover, for a much wider temperature regime (up to $t\sim 1$) than
predicted in Ref.~\cite{Barash:97}.  With respect to this discrepancy we note
that the faceting and the related angle averaging present in GBJs may have to
be taken into account in order to explain the measured temperature dependence
of the zero bias conductance peak.

%%%%%%%%%%%%%%%%%%%%%%%%  FIGURE 9 %%%%%%%%%%%%%%%%%%%%%%%%%

%%%%%%%%%%%%%%%%%%%%%%%%%%%%%%%%%%%%%%%%%%%%%%%%%%%%%%%%%%%%

Next we discuss the magnetic field dependence of the zero bias conductance
peak. In a recent paper Hu discusses the magnetic field effects that are
expected for GBJs including the possibility that the magnetic field might be
shielded or screened \cite{Hu:98}. However, shielding effects can be
definitely ruled out by our experimental results.  By studying the magnetic
field dependence of the critical current of symmetric HTS-GBJs, we always
find a Fraunhofer diffraction pattern like dependencies. From the distance of
the minima in this dependence we can conclude that the magnetic field
penetrates the GBJ as expected. Furthermore, the magnetic field penetrating
the grain boundary is enhanced by a factor of order $W/d$ due to flux
focusing effects \cite{Rosenthal:91}. Here, $W$ is the width of the GBJ and
$d$ the thickness of the junction electrodes.

We now discuss the fact that no direct splitting of the zero bias conductance
peak  in an applied magnetic field is observed in the GBJ experiments, whereas
such splitting is expected within the Andreev bound state model.  In agreement
to our observation no direct splitting of the zero bias conductance peak  has
been found also by Ekin {\em et al.} in experiments using planar SIN-type
junctions \cite{Ekin:97}. Furthermore, in experiments using low temperature
scanning tunneling spectroscopy a shift or a broadening of the zero bias
conductance peak  has been observed \cite{Kashiwaya:94}. Only in some cases a
clear split of the zero bias conductance peak  in an applied magnetic field
including a split at zero applied field was found
\cite{Covington:97,Lesueur:92}. Hu proposed to explain these different
observations with a different behaviour of the magnetic field penetration
\cite{Hu:98}. However, since the penetration of the magnetic field and flux
focusing effects are at least for HTS-GBJs well understood, this cannot be the
reason for the absence of a direct splitting of the zero bias conductance peak
for GBJs. However, this feature may be understood by considering the faceting
of the grain boundary or, equivalently, the interface roughness present in the
different junction types. Due to the faceting of the grain boundary barrier
the GBJ is known to be formed by a parallel array of many small junctions on a
micron and submicron scale with orientations that differ considerably from the
nominal misorientation angle of the grain boundary (see Fig.~\ref{gbj}).
Together with impurity scattering this may lead to a masking of the direct
splitting that is expected for junctions with perfectly flat interfaces in an
applied field \cite{Fogelstroem:98}.

\begin{figure}[htb]
\noindent
\vspace*{0cm}\\
\hspace*{0cm}\centering\epsfxsize=12.5cm\epsffile{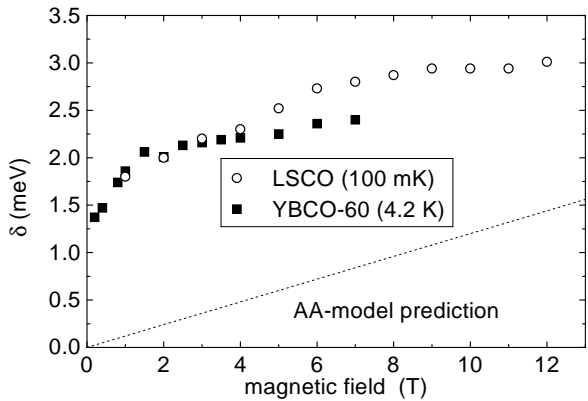}
\vspace*{-3.5cm}
\caption{Splitting $\delta$ of the $G(V,H)-G(V,H=0)$ curves of a 36.8$^{o}$ [001] tilt
LSCO-GBJ at $T=100$\,mK and a 24$^{o}$ [001] tilt YBCO-GBJ at $T=4.2$\,K
vs.~applied magnetic field. Also shown is the Appelbaum-Anderson-model
prediction for $g=2$.}
\label{epj98f8}
\end{figure}

Although the GBJ experiments show no direct splitting of the zero bias
conductance peak  in an applied magnetic field, there is a shift of spectral
weight from zero to finite energies.  This shift of spectral weight results in
two peaks in the $G(V,H)-G(V,0)$ curves. Analyzing these curves we can define
a splitting $\delta$ as half the peak-to-peak separation. In
Fig.~\ref{epj98f8}, the $\delta$ values derived in this way are plotted versus
$H$ for a LSCO-GBJ at 100\,mK and for a YBCO-GBJ at 4.2\,K. Clearly, $\delta$
does not vary linearly with the applied field as predicted by the
Appelbaum-Anderson model. For all investigated samples $\delta$ increases
linearly only at small fields below about 2\,T. However, for larger fields
$\delta$ clearly deviates from a linear behaviour and tends to saturate at a
constant value at high fields. This is in qualitative agreement with
experimental results published very recently \cite{Covington:97,Lesueur:92}
and theoretical predictions by Fogelstr\"{o}m {\it et al.}
\cite{Fogelstroem:97a}. We note that the splitting $\delta$ by definition has
to pass through the origin in Fig.~\ref{epj98f8}. That is, from our data we
cannot conclude that there is a finite splitting at zero applied magnetic
field as observed by Covington {\it et al.} \cite{Covington:97}, although this
is suggested by the data shown in Fig.~\ref{epj98f8}. The steep jump from zero
to a finite splitting for small applied magnetic field is characteristic for
all our samples. Furthermore, the magnetic field dependence of the indirect
splitting derived from the $G(V,H)-G(V,0)$ curves of GBJs (see
Fig.~\ref{epj98f8}) is consistent with the direct splitting observed by
Covington {\it et al.} \cite{Covington:97} for SIN-type junctions. The reason
for this consistency is not obvious and still has to be clarified.

%%%%%%%%%%%%%%%%%%%%%%%%  FIGURE 10 %%%%%%%%%%%%%%%%%%%%%%%%%

%%%%%%%%%%%%%%%%%%%%%%%%%%%%%%%%%%%%%%%%%%%%%%%%%%%%%%%%%%%%

We finally discuss a possible zero field splitting of the zero bias
conductance peak. As discussed above, a broken time reversal symmetry state
at the surface of a HTS can lead to a split of the zero bias conductance peak
in zero magnetic field. Such direct splitting has not been observed in our
experiments down to temperatures of 100\,mK for LSCO and down to 4.2\,K for
YBCO. Our observations are in agreement to the recent findings of other
groups \cite{Alff:96b,Ekin:97}. Although the detailed reason for the presence
or absence of a zero field splitting of the ZBCO is not clear, one can
speculate that the junction property preventing the observation of a direct
splitting of the zero bias conductance peak  by an applied magnetic field
also may result in the absence of any direct splitting in zero applied field.
The most likely reason is the roughness of the surface of the junction
electrodes, which in the case of GBJs is caused by the faceting of the grain
boundary. Surface roughness together with impurity scattering may lead to a
reduction of those quasiparticle trajectories that are responsible for the
splitting of the zero bias conductance peak. Furthermore, surface currents
along a strongly faceted GBJ could tend to cancel each other. Certainly, this
topic has to be clarified in more detail by future theoretical and
experimental work. Summarizing our discussion we note that at present it is
impossible to make a definitive statement with respect to the occurence of a
state with broken time reversal symmetry in HTS-GBJs.

\subsection{NCCO-spectra}
\label{ncco}

In the tunneling spectra of NCCO-GBJs never a zero bias conductance peak  has
been observed. This observation is consistent with an $s$-wave symmetry of
the order parameter in NCCO in contrast to other HTS such as YBCO, BSCCO, or
LSCO.  Since there is additional experimental evidence that NCCO has a
dominant $s$-wave component of the order parameter, the question arises
whether or not NCCO can be viewed as a classical BCS-superconductor.
Furthermore, experiments performed on NCCO are highly important in order to
clarify the origin of the specific features observed for the $d$-wave HTS. In
this subsection we discuss the tunneling spectra measured for NCCO-GBJs and
compare them to those obtained for the $d$-wave HTS as well as to those
expected for classical BCS-superconductors.

Discussing the NCCO spectra shown in Figs.~\ref{epj98f4} and~\ref{epj98f6},
one might think at first glance that the shape of the spectra resembles that
of a typical spectrum of a BCS superconductor, if a finite lifetime of the
quasiparticles is taken into account by introducing a smearing parameter
$\Gamma$ according to Dynes {\it et al.} \cite{Dynes:78}. However, looking
more closely to the spectra it becomes evident that the measured $G(V)$
curves cannot be described by a BCS type behaviour because of the large
conductance at zero bias and the V-shaped gap structure in contrast to the
U-shaped structure expected for conventional BCS superconductors. The
$G(V)$-curves of NCCO-GBJs resemble by far more those of GBJs fabricated from
the other ($d$-wave) HTS materials. We note that a fit of the measured
tunneling spectra using the Dynes formula \cite{Dynes:78} is possible only in
a few cases. Moreover, to obtain a reasonable fit $\Gamma\sim\Delta_0$ has to
be used what is difficult to justify. However, the Dynes expression cannot
account for both the sharp peaks at the gap edge and the high spectral weight
within the gap. Above we argued that NCCO is not a $d$-wave superconductor
because a zero bias conductance peak  is completely absent in the tunneling
spectra. To account for the observed features an anisotropic $s$-wave order
parameter can be assumed. Then, qualitatively the high spectral weight at
zero energy and the absence of the zero bias conductance peak  can be
explained. We finally note that an extended $s$-wave symmetry of the order
parameter cannot be fully ruled out at present, since the detailed influence
of the surface geometry on the tunneling spectra is not known.

\section{Conclusions}
\label{conc}

Tunneling spectroscopy on differently oriented surfaces of HTS can be used to
study the symmetry of the order parameter, since the tunneling spectra are
sensitive to the phase of the pair potential. Our comprehensive study of
various HTS-GBJs gives clear evidence for the presence of Andreev bound states
at the surface of the junction electrodes. For the materials YBCO, BSCCO, and
LSCO always a zero bias conductance peak  due to Andreev bound states was
observed what is consistent with a $d$-wave symmetry of the order parameter in
these materials. In contrast, NCCO never showed a zero bias conductance peak
what is consistent with the absence of Andreev bound states and an anisotropic
$s$-wave symmetry of the order parameter.  Our results show a clear
correlation between the presence of a zero bias conductance peak  and
superconductivity in the junction electrodes. This gives strong evidence for
an explanation of the zero bias conductance peak  in terms of Andreev bound
states and against an explanation in terms of magnetic impurity scattering.
The presence of Andreev bound states at surfaces and interfaces of $d$-wave
HTS also explains the observed anomalous temperature dependence of the London
penetration depth. Considering all tunneling data that is available at present
from experiments using different tunneling configurations, the vast majority
of the measured tunneling spectra can be well explained by taking into account
the presence of Andreev bound states at the surface of HTS due to a $d$-wave
symmetry of the order parameter. However, some issues such as the presence of
a broken time reversal symmetry state at surfaces of $d$-wave HTS are still
unsettled. In contrast to SIN-junctions, for HTS GBJs no direct splitting of
the zero bias conductance peak  in zero applied magnetic field was observed.

\section{Acknowledgements}

The authors thank A.~Beck, W.~Belzig, H.~Burkhardt, M.~Covington,
M.~Fogelstr\"{o}m, J.~Halbritter, S.~Kashiwaya, D.~Rainer, P.~Richter,
J.~Sauls, S.~Scheidl, and Y.~Tanaka for stimulating discussions. We are
grateful to R.~Dittmann, G.~Koren, M.~Naito, and H.~Sato for supplying
high-quality samples.

This work is supported by the Deutsche Forschungsgemeinschaft (SFB 341).

\end{document}